\documentclass[]{aastex631}

\usepackage[figuresleft]{rotating}
\usepackage{threeparttable}

\shorttitle{The Unusual AGN Host NGC~1266}
\shortauthors{Chen et al.}
\graphicspath{{./}{figures/}}

\begin{document}

\title{The Unusual AGN Host NGC~1266: Evidence for Shocks in a Molecular Gas Rich S0 Galaxy with a Low Luminosity Nucleus}

\author{Peibin Chen}
\affiliation{Department of Astronomy, Xiamen University, Xiamen, Fujian 361005, China; \textcolor{blue}{\emph{jfwang@xmu.edu.cn}}}

\author{Yinghe Zhao}
\affiliation{Yunnan Observatories, Chinese Academy of Sciences, Kunming 650011, China; \textcolor{blue}{\emph{zhaoyinghe@ynao.ac.cn}}}


\author{Junfeng Wang}
\affiliation{Department of Astronomy, Xiamen University, Xiamen, Fujian 361005, China; \textcolor{blue}{\emph{jfwang@xmu.edu.cn}}}



\begin{abstract}

NGC 1266 is a lenticular galaxy (S0) hosting an active galactic nucleus (AGN), and known to contain a large amount of shocked gas. We compare the luminosity ratio of mid-\emph{J} CO lines to IR continuum with star-forming galaxies (SFGs), and then model the CO spectral line energy distribution (SLED). We confirm that in the mid- and high-\emph{J} regions ($J_{\rm up}$ = 4--13), the C-type shock ($v_{\rm s}$ = 25 km s$^{-1}$, $n_{\rm H}$ = $5\times10^{4}$ cm$^{-3}$) can reproduce the CO observations well. The galaxy spectral energy distribution (SED) is constructed and modeled by the code {\tt X-CIGALE} and obtains a set of physical parameters including the star formation rate (SFR, 1.17 $\pm$ 0.47 \emph{M$_{\odot}$}yr$^{-1}$). Also, our work provides SFR derivation of [C\,{\sc ii}] from the neutral hydrogen regions only (1.38 $\pm$ 0.14 $M_{\odot}$yr$^{-1}$). Previous studies have illusive conclusions on the AGN or starburst nature of the NGC 1266 nucleus. Our SED model shows that the hidden AGN in the system is intrinsically low-luminosity, consequently the infrared luminosity of the AGN does not reach the expected level. Archival data from \emph{NuSTAR} hard X-ray observations in the 3--79 keV band shows a marginal detection, disfavoring presence of an obscured luminous AGN and implying that a compact starburst is more likely dominant for the NGC 1266 nucleus.

\end{abstract}

\keywords{galaxies: elliptical and lenticular, cD - galaxies: active - galaxies: individual (NGC 1266) - galaxies: ISM - galaxies: star formation}

Accepted by Research in Astronomy and Astrophysics
\section{Introduction} \label{sec:intro}

The morphology of galaxies plays an key role in the galaxy formation and evolution (e.g., morphological quenching; \citealt{2009ApJ...707..250M}). The Hubble tuning-fork diagram divides galaxies into early-type galaxies (ETGs) and late-type galaxies (LTGs), and in between is their bridge--S0 galaxies. S0 galaxies appear as a central bulge surrounded by a disk, but they do not show the spiral patterns \citep{1970ApJ...160..811F, 1978ApJ...225..742S}. Its bulge is almost as large as the disk, so the galaxy appears almost spherical in shape. ETGs generally contain all types of ellipticals and S0 galaxies (e.g., \citealt{2022MNRAS.509.1237X}). It is believed that they are short of atomic and molecular gas \citep{1991ApJ...379..177L}, but their disk galaxy progenitors are generally gas-rich \citep{2003MNRAS.341...33K}. The conversion of gas to stars in ETGs is related to a series of key processes in galaxy evolution, e.g., starburst and AGN activities. 

ETGs are generally redder than LTGs \citep{1978ApJ...225..742S}, and contain little or no ongoing star formation (SF). Nevertheless, atomic gas can exist in ETGs, and dust and molecular gas have been observed as well (e.g., \citealt{2001AJ....121..808C, 2007MNRAS.377.1795C}). Their related SF activities have been revealed through the continuum \citep{2008AJ....136.2782L} or the emission lines diagnosis (e.g., \citealt{2015ApJ...798...31A, 2016ApJ...831...63X}). Some ETGs have higher gas surface density ($\Sigma_{\rm gas}$) than normal spirals \citep{2014MNRAS.444.3427D}, yet their SFR surface density ($\Sigma_{\rm SFR}$) is significantly lower than expected by the Kennicutt-Schmidt law \citep{1959ApJ...129..243S, 1998ApJ...498..541K}. Moreover, studies have shown that different environments may have various effects on the formation and evolution of S0 galaxies (e.g., \citealt{2021MNRAS.508..895D, 2022MNRAS.509.1237X}). Therefore, the studies of S0 galaxies not only help to understand the quenching or rejuvenation of SF activities in ETGs, but also can shed light on the nature of the SF law in ETGs (e.g., \citealt{2022MNRAS.509.1237X}).

The nearby S0 galaxy NGC 1266 has a luminosity distance of $D_{\rm L}$ = 30.6 Mpc (\emph{z} = 0.0072, \citealt{2009ApJ...707..890T}). It contains 1.1$\times$10$^9$ \mit M$_{\odot}$ molecular gas within a radius of 60 pc \citep{2011ApJ...735...88A}, a significant fraction of which is outflowing. As one of the ETGs that have been mapped interferometrically\footnote{Based on millimeter interferometric IRAM PdBI and optical integral field unit SAURON observations.} ($\sim$ 20 galaxies, \citealt{2011MNRAS.410.1197C}), the molecular gas in NGC 1266 is remarkable due to its compactness and the massive outflow. For comparison, previous work found the existence of a small sample of galaxies (e.g., M82, 3C 293, Mrk 231) with molecular outflows (e.g., \citealt{2002ApJ...580L..21W, 2005MNRAS.362..931E, 2010A&A...518L.155F}). The molecular outflows were thought to be triggered by mergers or interactions. Compared to this sample, it is intriguing that the isolated galaxy NGC 1266 does not show any signs of interaction or merger activity (e.g., \citealt{2011ApJ...735...88A}).

A series of studies are reported in the literature for NGC 1266, including its molecular gas properties (e.g., \citealt{2011ApJ...735...88A, 2015ApJ...800..105G}), SF processes (e.g., \citealt{2014A&A...568A..62D, 2015ApJ...798...31A}) and gas excitation mechanisms (e.g., \citealt{2012MNRAS.426.1574D, 2013ApJ...779L..19P}). Among these, \citet{2011ApJ...735...88A} confirmed that the existing molecular outflow is driven by the mechanical energy of the AGN. The derived mass loss rate (110 \emph{M$_{\odot}$}~yr$^{-1}$) was at least 100 times that of the SFR \citep{2015ApJ...798...31A}. The derivation of SFR, however, has many limitations in this system. This is because of the presence of AGN in NGC 1266, and it also contains a large amount of shocked gas \citep{2012MNRAS.426.1574D}. \citet{2015ApJ...798...31A} obtained the SFR through the continuum or the ionized gas emission lines and they suggested that 0.3 \emph{M$_{\odot}$}yr$^{-1}$ (polycyclic aromatic hydrocarbons, PAHs + UV) is the lower limit of SFR. \citet{2012MNRAS.426.1574D} used different methods (line ratio diagnostics and large-scale velocity structure) to prove that the ionized gas emission line ratio in the system is consistent with the result of the shock ionization. \citet{2013ApJ...779L..19P} and \citet{2015ApJ...800..105G} analyzed the CO transitions with the existing models and obtained evidence that the shock dominates the gas excitation. Moreover, the dense gas observations (e.g., CS(2-1) and HCN(1-0), \citealt{2015ApJ...798...31A}) confirmed that the line-of-sight gas column density towards the AGN is Compton-thick (\citealt{2015ApJ...798...31A}, $N_{\rm H} \sim 10^{24} {\rm cm}^{-2}$). These previous studies, however, are inconclusive on the nature of activity in the center of NGC 1266 (a radiatively powerful AGN or an ultra-compact starburst, e.g., \citealt{2014A&A...568A..62D, 2015ApJ...798...31A}). 

In this work, we present the analysis of infrared (IR) and sub-millimeter (sub-mm) observations of NGC 1266, investigating its SFR, the CO spectral line energy distribution (SLED) and the nature of its nucleus. This paper is organized as follows. In section \ref{section2}, we describe the archival data in this paper, as well as the modeling process. In section \ref{section3}, we discuss the diagnosis of the energy sources and the SFR derivation. Consequently, based on the model fitting of CO SLED, we confirm that shock dominates the mid- to high-\emph{J} CO excitation ($J_{\rm up}$ = 4 -- 13). We further validate this result by calculating the luminosity ratio of mid-\emph{J} CO lines ($J_{\rm up}$ = 4 -- 10) to IR continuum, $R_{\rm mid\ CO}$ value \citep{2014ApJ...787L..23L}. Finally, the SED of the galaxy is constructed and modeled by the code {\tt X-CIGALE}. In addition, this work also compares other commonly used methods (e.g., IR continuum, [C\,{\sc ii}]) for SFR derivation. In sections \ref{section4} and \ref{section5}, we discuss the activity within NGC 1266 and summarize our work. In the Appendix \ref{appA}, we provide the SED fitting parameter settings.

\section{Archival DATA AND Modeling Process}
\label{section2}
\subsection{Archival Data}
\emph{Herschel} Spectral and Photometric Imaging Receiver (SPIRE) and Photodetector Array Camera and Spectrometer (PACS) observations of NGC 1266 are collected. We obtain the mid- and high-\emph{J} CO data from the program \emph{Beyond The Peak} (BTP; OT1$\_$jsimth; PI: J. D. Smith), which conducted a survey on 21 nuclear regions and 2 extra-nuclear regions from the \emph{SINGS} \citep{2003PASP..115..928K} and \emph{KINGFISH} \citep{2011PASP..123.1347K}) surveys. It completed deep, intermediate-spaced mapping of transitions above CO(3-2) in these nearby galaxies. Also, as part of the \emph{Herschel} key program \emph{KINGFISH}, NGC 1266 has been imaged in the emission lines (e.g., [C\,{\sc ii}], [N\,{\sc ii}] and [O\,{\sc iii}]). The data analysis is previously presented in \citet{2012ApJ...747...81C}. Our CO data ($J_{\rm up}$ = 4 -- 13) come from \citet{2013ApJ...779L..19P} (listed in Table \ref{Table1}), and the ionized emission lines data ([C\,{\sc ii}] and [N\,{\sc ii}] 205 $\rm{\mu m}$) come from \citet{2019ApJ...886...60S}.

We further obtain archival CO data ($J_{\rm up}$ = 1 -- 3) observed with \emph{IRAM} 30 m telescope and Sub-millimeter Array (\emph{SMA}). NGC 1266 is part of the CO(1-0), (2-1) survey of a flux-limited sample of all ATLAS$^{\rm 3D}$ galaxies \citep{2011MNRAS.414..940Y}, observed with \emph{IRAM} 30 m telescope. The data consist of a single pointing to the galaxy center, covering a velocity range of 1300 km s$^{-1}$ for the CO spectra. The spectral resolutions for CO(1-0) and CO(2-1) are 2.6 km s$^{-1}$ and 5.2 km s$^{-1}$, respectively, with 2155 km s$^{-1}$ as the velocity center and 4.73 Jy K$^{-1}$ as the conversion factor \citep{2011ApJ...735...88A}. In addition, the CO(3-2) was observed by \emph{SMA} at 345 GHz, with an imaging bandwidth of 2600 km s$^{-1}$ and a spectral resolution of 0.7 km s$^{-1}$ (centered on the system velocity). A detailed description of the data can be found in \citet{2011ApJ...735...88A}. Our lowest-$J$ (1-0,2-1,3-2) CO transitions listed in Table \ref{Table1} are adopted from \citet{2011ApJ...735...88A}.

Finally, we also utilized unpublished X-ray data from \emph{NuSTAR} \citep{2016AAS...22724356L}. If the gas column density is indeed as high as $6 \times 10^{24}$ cm$^{-2}$ \citep{2015ApJ...798...31A}, the hidden Compton-thick AGN could still be detected by hard X-ray observations that are sensitive to $>10$~keV.  We used the 3--79 keV archival data from \emph{NuSTAR} hard X-ray observation (obsID 60101048002 with 51.5 ks exposure time, PI: L. Lanz) to constrain this. We run the NuSTAR Data Analysis Software (NuSTARDAS v2.0.0, \citealt{2020arXiv200500569M}) with the updated calibration files (CALDB version 20220413) to obtain pipeline processed data. Our analysis shows that the detected counts on both Focal Plane Module A and B (FPMA and FPMB) detectors give a net count rate $1.8\pm 0.6 \times 10^{-3}$, indicating a marginal 3$\sigma$ detection. If we accept the large column density of $6 \times 10^{24}$ cm$^{-2}$ from dense molecular gas confirmed by \cite{2015ApJ...798...31A}, using the PIMMS\footnote{\tiny{\url{https://heasarc.gsfc.nasa.gov/cgi-bin/Tools/w3pimms/w3pimms.pl}}} tool and adopting a power law continuum with a typical photon index $\Gamma=1.8$, the estimated intrinsic 2--10 keV luminosity is $4.50 \times 10^{40}$ erg s$^{-1}$. In addition, we adopt the multi-wavelength photometric data from the \emph{NASA/IPAC Extragalactic Database}\footnote{\tiny{\url{https://ned.ipac.caltech.edu}}} (NED, listed in Table \ref{Table2}) to construct the SED of NGC 1266.

\begin{table}
\begin{center}
\caption{The measurements of the CO line flux in NGC 1266}\label{Table1}
\begin{threeparttable}
 \begin{tabular}{ccccc}
  \hline\hline\noalign{\smallskip}
Lines\tnote{a} &  Trans      & Wave ($\rm{\mu m}$) & \begin{tabular}[c]{@{}l@{}}Flux\\ (10$^{-18}$ w m$^{-2}$)\end{tabular} & Telescope (Beam size)                    \\
  \hline\noalign{\smallskip}
$^{12}{\rm C}{\rm O}$  & 1--0 & 2600.80     & 0.63 ($\pm$ 0.04)  &  $IRAM$ 30 m (21.6$^{\prime \prime}$)  \\ 
$^{12}{\rm C}{\rm O}$  & 2--1  & 1300.40     & 3.90 ($\pm$ 0.15)  &  $IRAM$ 30 m (12$^{\prime \prime}$)                   \\
$^{12}{\rm C}{\rm O}$  & 3--2  & 867.00      & 10.20 ($\pm$ 1.10)  &  $SMA$ (30$^{\prime \prime}$)                  \\
$^{12}{\rm C}{\rm O}$  & 4--3  & 650.70      & 26.90 ($\pm$ 11.40)  &  $Herschel$ (42.8$^{\prime \prime}$)                  \\
$^{12}{\rm C}{\rm O}$  & 5--4  & 520.60      & 38.40 ($\pm$ 6.90)  &  $Herschel$ (35.2$^{\prime \prime}$)                  \\
$^{12}{\rm C}{\rm O}$  & 6--5  & 433.90      & 42.80 ($\pm$ 3.50)  &  $Herschel$ (31.2$^{\prime \prime}$)                  \\
$^{12}{\rm C}{\rm O}$  & 7--6  & 371.90      &47.60 ($\pm$ 2.80)  &  $Herschel$ (35.9$^{\prime \prime}$)                  \\
$^{12}{\rm C}{\rm O}$  & 8--7  & 325.50      & 47.10 ($\pm$ 3.60)  &  $Herschel$ (40.1$^{\prime \prime}$)                  \\
$^{12}{\rm C}{\rm O}$  & 9--8  & 289.30      & 46.40 ($\pm$ 6.50)  &  $Herschel$ (19.0$^{\prime \prime}$)                  \\
$^{12}{\rm C}{\rm O}$  & 10--9  & 260.40      & 39.40 ($\pm$ 6.00)  &  $Herschel$ (17.4$^{\prime \prime}$)                  \\
$^{12}{\rm C}{\rm O}$  & 11--10  & 236.80      & 40.30 ($\pm$ 6.10)  &  $Herschel$ (17.3$^{\prime \prime}$)                  \\
$^{12}{\rm C}{\rm O}$  & 12--10  & 217.10      & 28.50 ($\pm$ 3.00)  &  $Herschel$ (16.9$^{\prime \prime}$)                  \\
$^{12}{\rm C}{\rm O}$  & 13--12  & 200.40      & 26.50 ($\pm$ 5.80)  &  $Herschel$ (16.6$^{\prime \prime}$)                  \\
  \noalign{\smallskip}\hline
\end{tabular}
\begin{tablenotes}
 \footnotesize
 \item[a] The ground CO lines (J$_{\rm up}$ = 1--3) are from \citet{2011ApJ...735...88A}, and the rest of CO (J$_{\rm up}$ = 4--13) is from \citet{2013ApJ...779L..19P} by \emph{Herschel} observations.
\end{tablenotes}
\end{threeparttable}
\end{center}
\end{table}

\begin{table}
\begin{center}
\tiny
\caption{The NGC 1266 Global Flux Densities in Janskys}\label{Table2}
\begin{threeparttable}
 \begin{tabular}{ccccccc}
  \hline\hline\noalign{\smallskip}
Filters &  \begin{tabular}[c]{@{}c@{}}$\lambda_{\rm eff}\tnote{a}$\\ ($\rm {\mu m}$)\end{tabular}      & \begin{tabular}[c]{@{}c@{}}Flux (observed)\\ (Jy)\end{tabular}  & \begin{tabular}[c]{@{}l@{}}Obs$\_$uncertainty\tnote{b} \\ (Jy)\end{tabular} & A$_{\lambda}$/A$_V$\tnote{c}  &  \begin{tabular}[c]{@{}c@{}}Flux (corrected)\\ (Jy)\end{tabular}  &  \begin{tabular}[c]{@{}l@{}}Corrected$\_$uncertainty\tnote{b} \\ (Jy)\end{tabular}                   \\
  \hline\noalign{\smallskip}
GALEX$\_$FUV        & 0.152                                                                                  & 4.90 $\times$ 10$^{-5}$                                            & 7.00 $\times$ 10$^{-6}$                                              & 0.78               & 0.10 $\times$ 10$^{-3}$ & 0.85 $\times$ 10$^{-6}$  \\ 
GALEX$\_$NUV        & 0.231                                                                                  & 2.90 $\times$ 10$^{-4}$                                            & 4.00 $\times$ 10$^{-5}$                                              & 0.90               & 0.63 $\times$ 10$^{-3}$ & 0.49 $\times$ 10$^{-5}$                   \\
B                & 0.460                                                                                  & 2.00 $\times$ 10$^{-2}$                                                              & 4.00 $\times$ 10$^{-4}$                                              & 0.39                & 3.00 $\times$ 10$^{-2}$ & 0.49 $\times$ 10$^{-4}$                  \\
V                & 0.538                                                                                  & 3.60 $\times$ 10$^{-2}$                                                             & 4.00 $\times$ 10$^{-4}$                                              & 0.305                 & 5.00 $\times$ 10$^{-2}$ & 0.49 $\times$ 10$^{-4}$                  \\
R                & 0.652                                                                                  & 3.70 $\times$ 10$^{-2}$                                                             & 4.00 $\times$ 10$^{-4}$                                              & 2.36 $\times$ 10$^{-1}$               & 5.00 $\times$ 10$^{-2}$ & 0.49 $\times$ 10$^{-4}$                  \\
I                & 0.866                                                                                  & 3.50 $\times$ 10$^{-2}$                                                             & 4.00 $\times$ 10$^{-4}$                                              & 1.73 $\times$ 10$^{-1}$               & 4.00 $\times$ 10$^{-2}$ & 0.49 $\times$ 10$^{-4}$                  \\
2MASS$\_$J          & 1.20                                                                                    & 1.20 $\times$ 10$^{-1}$                                                              & 4.40 $\times$ 10$^{-4}$                                              & 0.89 $\times$ 10$^{-1}$               & 1.30 $\times$ 10$^{-1}$ & 0.53 $\times$ 10$^{-4}$                  \\
2MASS$\_$H          & 1.67                                                                                   & 1.30 $\times$ 10$^{-1}$                                                              & 6.14 $\times$ 10$^{-4}$                                             & 0.56 $\times$ 10$^{-1}$               & 1.40 $\times$ 10$^{-1}$ & 0.75 $\times$ 10$^{-4}$                  \\
2MASS$\_$Ks         & 2.20                                                                                    & 1.20 $\times$ 10$^{-1}$                                                              & 5.86 $\times$ 10$^{-4}$                                             & 0.35 $\times$ 10$^{-1}$               & 1.20 $\times$ 10$^{-1}$ & 1.26 $\times$ 10$^{-4}$                  \\
IRAC1            & 3.56                                                                                   & 5.50 $\times$ 10$^{-2}$                                            & 8.00 $\times$ 10$^{-3}$                                              & 1.35 $\times$ 10$^{-2}$              & 0.60 $\times$ 10$^{-1}$ & 0.97 $\times$ 10$^{-3}$                  \\
IRAC2            & 4.51                                                                                   & 4.20 $\times$ 10$^{-2}$                                            & 6.00 $\times$ 10$^{-3}$                                              & 0.86 $\times$ 10$^{-2}$              & 0.40 $\times$ 10$^{-1}$ & 0.73 $\times$ 10$^{-3}$                  \\
IRAC3            & 5.76                                                                                   & 5.70 $\times$ 10$^{-2}$                                            & 8.00 $\times$ 10$^{-3}$                                              & 0.58 $\times$ 10$^{-2}$              & 0.60 $\times$ 10$^{-1}$ & 0.97 $\times$ 10$^{-3}$                  \\
IRAC4            & 7.69                                                                                   & 9.00 $\times$ 10$^{-2}$                                            & 1.20 $\times$ 10$^{-2}$                                              & 0.89 $\times$ 10$^{-2}$              & 0.90 $\times$ 10$^{-1}$ & 0.15 $\times$ 10$^{-2}$                  \\
WISE$\_$W3          & 12.8                                                                                   & 9.31 $\times$ 10$^{-2}$                                           & 5.15 $\times$ 10$^{-4}$                                             & 1.07 $\times$ 10$^{-2}$              & 0.90 $\times$ 10$^{-1}$ & 0.63 $\times$ 10$^{-4}$                  \\
MIPS$\_$24 & 23.80                                                                                   & 8.80 $\times$ 10$^{-1}$                                                              & 4.00 $\times$ 10$^{-2}$                                              & 0.58 $\times$ 10$^{-2}$              & 8.80 $\times$ 10$^{-1}$ & 0.49 $\times$ 10$^{-2}$                  \\
MIPS$\_$70\tnote{d}          & 70                                                                                     & 12.70                                                              &  0.95                                             & 0.69 $\times$ 10$^{-3}$            & 12.7 &   0.12                \\
PACS$\_$70\tnote{d}          & 72                                                                                     & 14.50                                                              & 0.70                                             & 0.69 $\times$ 10$^{-3}$             & 14.50 & 0.09                  \\
PACS$\_$100         & 103                                                                                    & 15.90                                                              &  0.80                                              & 0.31 $\times$ 10$^{-3}$             & 15.90 &  0.10                 \\
MIPS$\_$160         & 159                                                                                    & 10.30                                                              &  0.13                                          & 0.12 $\times$ 10$^{-3}$          & 10.30 &  0.02                  \\
PACS$\_$160         & 160                                                                                    & 11.30                                                              &  0.60                                              & 0.12 $\times$ 10$^{-3}$              & 11.30 &  0.07                  \\
SPIRE$\_$250        & 250                                                                                    & 3.61                                                              &  0.54                                             & 0.60 $\times$ 10$^{-4}$              & 3.61 &  0.07                  \\
SPIRE$\_$350        & 350                                                                                    & 1.48                                                              &  0.23                                             & 0.21 $\times$ 10$^{-4}$             & 1.48 &  0.03                  \\
SPIRE$\_$500        & 500                                                                                    & 4.15 $\times$ 10$^{-1}$                                                             & 7.03 $\times$ 10$^{-2}$                                             & 0.12 $\times$ 10$^{-4}$              & 4.20 $\times$ 10$^{-1}$ & 0.85 $\times$ 10$^{-2}$                 \\
  \noalign{\smallskip}\hline
\end{tabular}
\begin{tablenotes}
 \footnotesize
 \item[a] The filter central wavelengths are computed via Equation (1) and (2) from \citet{2017ApJ...837...90D}, and they are arranged in ascending order.
 \item[b] The uncertainties include statistical and systematic effects. 
 \item[c] Assuming an extinction to reddening ratio A$_{\rm V}$/E(B-V) = 3.1, E(B-V) = 0.098 mag from \citet{1998ApJ...500..525S}.
  \item[d] Shifting this effective wavelength to the modified value given by \citet{2014PASA...31...49B}.
\end{tablenotes}
\end{threeparttable}
\end{center}
\end{table}

\subsection{Modeling Process}
\subsubsection{CO SLED Fitting}
To diagnose the emission mechanism of the dominant spectral line in NGC 1266, we combined with previous studies \citep{2015ApJ...800..105G, 2015A&A...578A..95I} and firstly used the non-local thermodynamic equilibrium (non--LTE) code {\tt RADEX} \citep{2007A&A...468..627V} to model the CO observations. There are many parameters in this model, here we only focus on the non-default ones: kinetic temperature ($T_{\rm kin}$), gas density ($n_{\rm H_{2}}$) and CO column density ($N_{\rm CO}$). The statistical equilibrium calculation performed by {\tt RADEX} includes collisional and radiative processes. For CO molecular data (\citealt{2005A&A...432..369S}\footnote{\tiny{\url{home.strw.leidenuniv.nl/moldata}}}) and the collisional rate coefficient with H$_{2}$, we follow \citet{2010ApJ...718.1062Y} and assume that H$_{2}$ is the only collision counterpart. In our run, these parameters ranged from: 20--40 K ($T_{\rm kin}$), 10$^3$--10$^5$ cm$^{-3}$ ($n_{\rm H_{2}}$), and 10$^{14}$--10$^{17}$ cm$^{-2}$ ($N_{\rm CO}$). The excitation temperature in the model should cover a wide range, but we again performed verification work on the basis of previous work \citep{2015ApJ...800..105G, 2015A&A...578A..95I}, so we narrowed this range. For a wide range of input parameters, the code is able to generate a grid of model CO SLED (e.g., \citealt{2015ApJ...800..105G}).

\begin{figure}[!hbtp]
 \centering
 \includegraphics[width=\textwidth]{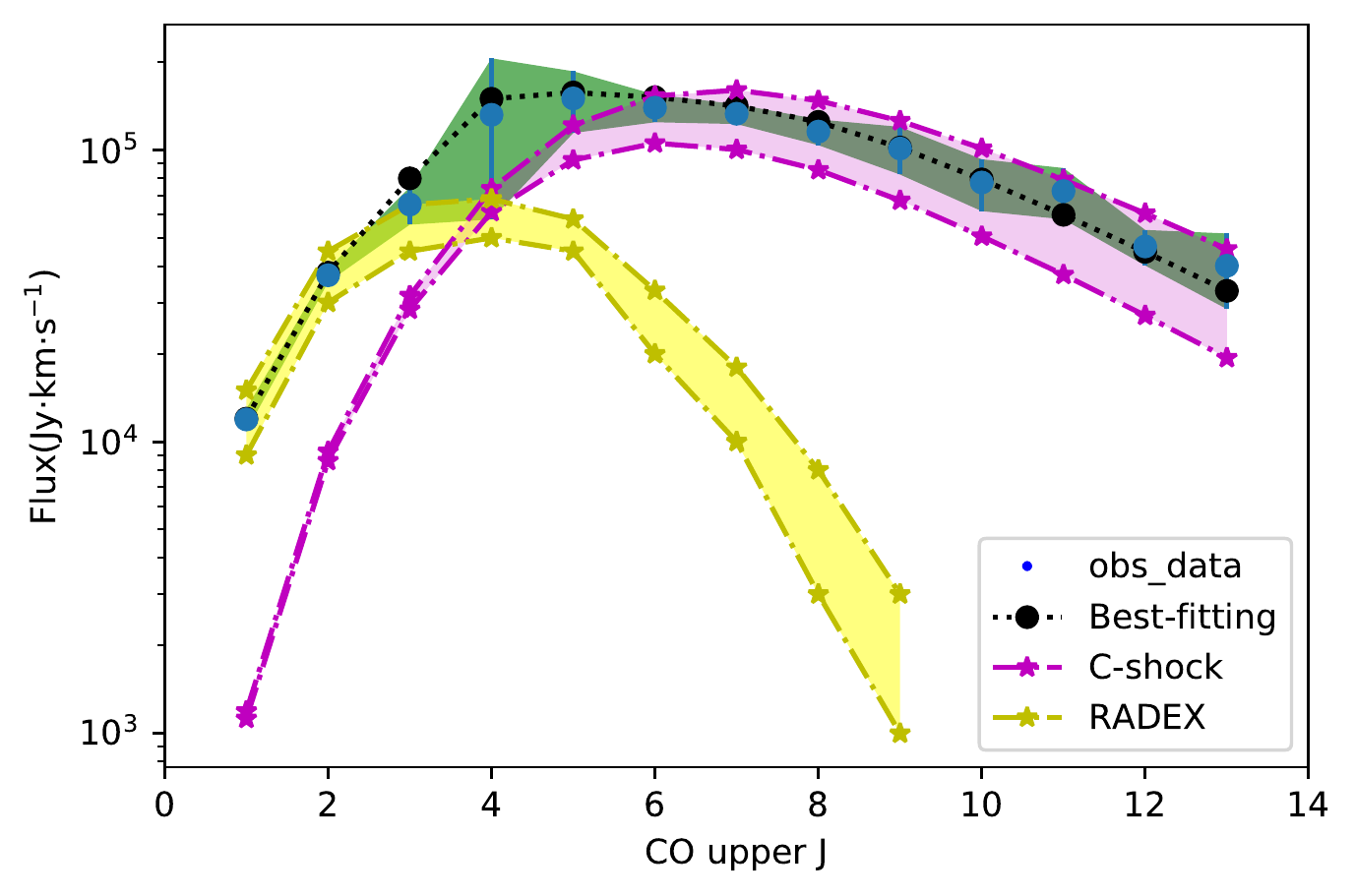}
  \caption{The observed CO SLED (blue circles with error bars) with the model fit. The best-fitting result is given by the black circles (reduced $\chi^2 \sim 2.0$). The parameters for the best fit are: $v_{\rm s}$ = 25 ($\pm$ 5) km s$^{-1}$, $n_{\rm H}$ = 5$\times$10$^4$ cm$^{-3}$ ({\tt shock} model); $T_{\rm kin}$ = 30 ($\pm$ 5) K, $n_{\rm H_2}$ = 1$\times$10$^4$ cm$^{-3}$ ({\tt RADEX} model). The shadow areas show the optimal solutions of the C-type shock models (magenta, the upper limit means $v_{\rm s}$ = 30 km s$^{-1}$, and the lower limit means $v_{\rm s}$ = 20 km s$^{-1}$, and $n_{\rm H}$ = 5$\times$10$^4$ cm$^{-3}$) and {\tt RADEX} models (yellow, the upper limit means $T_{\rm kin}$ = 35 K, and the lower limit means $T_{\rm kin}$ = 25 K, and $n_{\rm H_2}$ = 1$\times$10$^4$ cm$^{-3}$) for observations. The observed CO lines flux is listed in Table \ref{Table1}.}
 \label{Fig.2}
 \end{figure}   

Secondly, due to the existence of shocked gas in NGC 1266, we combined with previous work \citep{2010A&A...518L..42V, 2012ApJ...758..108S} and considered using the {\tt shock} code \citep{2015A&A...578A..63F}. The code has a number of input files, including files that specify initial values for dynamical and chemical variables, chemical reaction networks, data for calculating atomic and molecular intensities, and variables that control the precision of the numerical integration process \citep{2015A&A...578A..63F}. Combining with previous work \citep{1983ApJ...264..485D, 2013A&A...550A.106L, 2013ApJ...762L..16M, 2015A&A...578A..63F}, parameters\footnote{The magnetic field intensity is related to the hydrogen number density, which is defined as: $B(\mu G) = b[n_{\rm H}(\rm cm^{-3})]^{1/2}$. The $n_{\rm H}$ is defined as $n_{\rm H} = n(\rm{H}) + 2n(\rm{H_2}) + n(\rm{H^+})$.} of this model are set in our run as: C-type shock; the number of the fluid is 3 (for general C-type shock); 1.0--5.0 (magnetic field intensity $B$ = 10$^{1.5}$ -- 5 $\times$ 10$^{7/2}$ $\mu G$); $10^3$--$10^7$ cm$^{-3}$ (density of hydrogen nuclei, $n_{\rm H}$); 10--50 km s$^{-1}$ (shock velocity, $v_{\rm s}$). The large velocity gradient (LVG) calculations are part of the model, which yield the density and emission per unit volume in various molecular lines \citep{2015A&A...578A..63F}.

The two models will produce a series of meshes in these parameter spaces. Using a chi-square ($\chi^2$) fit between the CO observations and these simulated values, we obtain the best fit of the observations (the black circles in Figure \ref{Fig.2}). The best-fitting parameters are: $v_{\rm s}$ = 25 ($\pm$5) km s$^{-1}$, $n_{\rm H}$ = 5$\times$10$^4$ cm$^{-3}$ ({\tt shock} model); $T_{\rm kin}$ = 30 ($\pm$ 5) K, $n_{\rm H_2}$ = 1$\times$10$^4$ cm$^{-3}$ ({\tt RADEX} model).

\subsubsection{SED Fitting}
We use the {\tt X-CIGALE} code \citep{2020MNRAS.491..740Y} to construct its SED, which is modified on the basis of the python code {\tt CIGALE} \citep{2019A&A...622A.103B}. In fact, the model is constructed from many components: SFH, single-age stellar populations (SSPs), ionized gas templates including lines and continuum, flexible dust decay curves, dust emission templates, synchrotron emission, and finally the intergalactic medium influences \citep{2019A&A...622A.103B}. Each component contains different modules and is independently computed by these modules. For example, dust attenuation can be modeled alternatively with the \citet{2000ApJ...539..718C} or the \citet{2000ApJ...533..682C} models. These photometric measurements were unified in the work of \citet{2017ApJ...837...90D}, so we refer to their work (see their Table 2). For Galactic extinction correction, we used the color excess E(B-V) = 0.098 mag from \citet{1998ApJ...500..525S}. Adopting the extinction to reddening ratio $A_{\rm V}$/E(B-V) = 3.1, the V-band extinction is $A_{\rm V}$ = 0.305 mag. Table~\ref{Table2} lists the flux after the Galactic extinction correction and the corresponding $A_{\lambda}$/$A_{\rm V}$ value of each band. In the Appendix \ref{appA}, we provide the SED fitting parameter settings.

\section{ANALYSIS}
\label{section3}
\begin{figure}[!hbtp]
   \centering
   \includegraphics[width=\textwidth, angle=0]{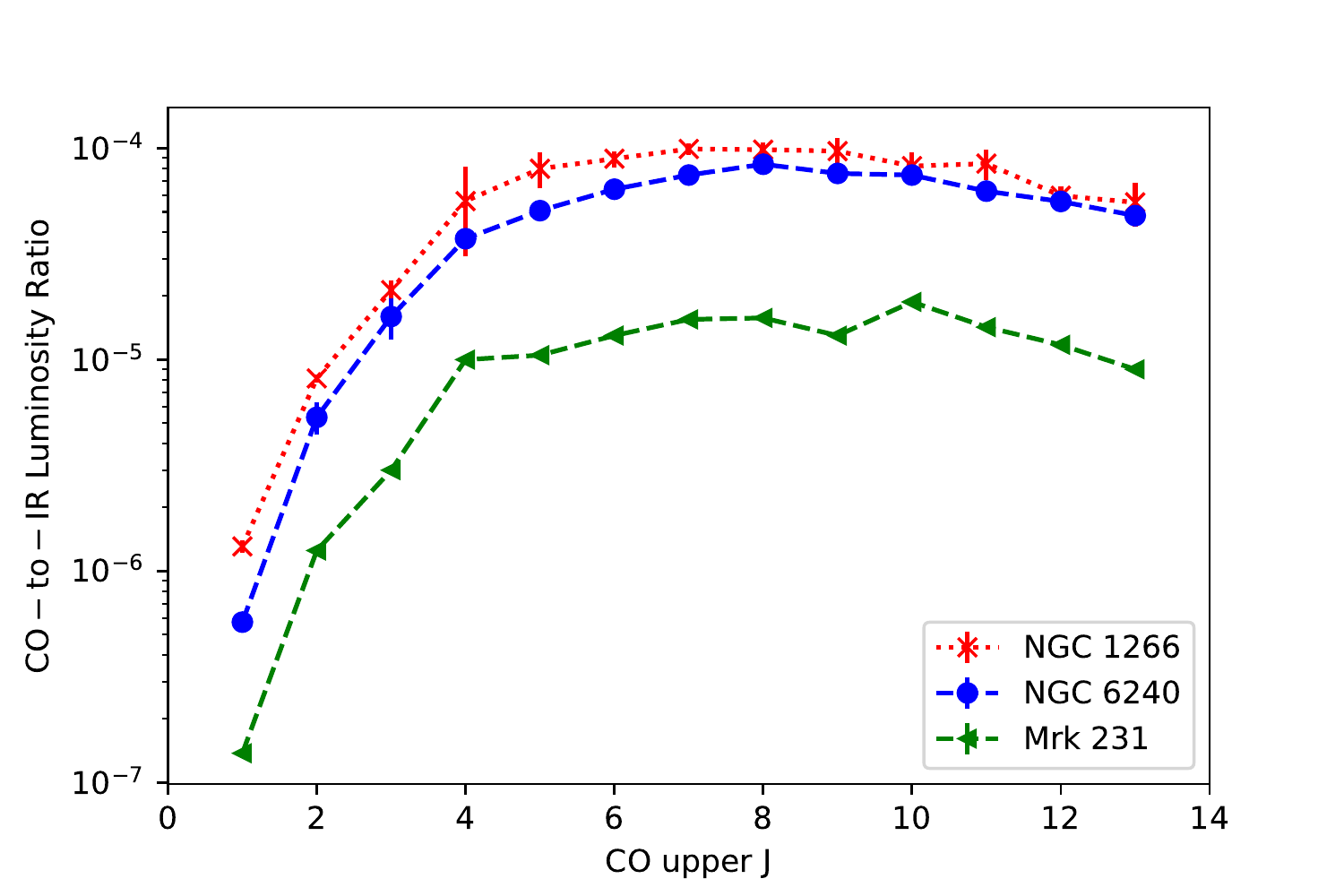}
   \caption{Comparison of the CO ladder of NGC 1266 (red crosses) to those obtained for NGC 6240 (blue circles, previous studied by \citealt{2013ApJ...762L..16M}) and Mrk 231 (green triangles, previous studied by \citealt{2010A&A...518L..42V}). The observed CO lines flux (listed in Table \ref{Table1}) is normalized by the corresponding $L_{\rm IR}$.}
   \label{Fig.1}
   \end{figure}

The shape of the CO ladder constructed by multiple CO transitions in NGC 1266 (Figure \ref{Fig.1}) is similar to that of the shock-dominant galaxy NGC 6240 (\citealt{2013ApJ...762L..16M}) and the AGN-dominant galaxy Mrk 231 (\citealt{2010A&A...518L..42V}), and the ratios of the CO luminosity ($J_{\rm up}$ = 1--13) to IR continuum ($L_{\rm CO}$/$L_{\rm IR}$\footnote{$L_{\rm CO}$ is the sum of all $J$ ladders from $J_{\rm up}$ = 1 to $J_{\rm up}$ = 13, $L_{\rm IR}$ is calculated by using the IRAS four-band fluxes, and the equation is $L_{\rm IR} = 4\pi D_{\rm L}^2 f_{\rm IR}$ from \citet{1996ARA&A..34..749S}, and the $f_{\rm IR} = 1.8\times10^{-14}(13.48f_{12} + 5.16f_{25} + 2.58f_{60} + f_{100}$ (w$\cdot$m$^{-2}$).}) are also very close (${L_{\rm CO}/L_{\rm IR}}_{\rm NGC1266}$ = 0.54$\times$10$^{-3}$, ${L_{\rm CO}/L_{\rm IR}}_{\rm NGC6240}$ = 0.67$\times$10$^{-3}$). Both are an order of magnitude larger than that of the AGN-dominant galaxy Mrk 231 (0.70$\times$10$^{-4}$). This suggests that it seems reasonable to assume that shocks dominate the gas excitation in NGC 1266. This comparative approach has been demonstrated in the study of NGC 6240 by \citet{2013ApJ...762L..16M}. Indeed, \citet{2013ApJ...779L..19P} and \citet{2015ApJ...800..105G} have confirmed the evidence of shock dominance under the comparisons of existing models ({\tt photo-dissociation regions(PDRs)}, {\tt RADEX} and {\tt shock}).

In this work, the heating mechanism of the galaxy is first analyzed using the ${\rm R_{mid\ CO}}$ value. Then the CO SLED is fitted with the help of existing models ({\tt RADEX}+{\tt shock}). The two methods yield consistent results, confirming previous work (e.g., \citealt{2013ApJ...779L..19P, 2015ApJ...800..105G}). Due to the difficulty of SFR derivation in NGC 1266, we used different methods to calculate its SFR (section \ref{section3.2}).

\subsection{Powering Source of the Molecular Line Emission}
\label{section3.1}
The molecular gas can be usually traced by its brightest CO lines \citep{1991ARA&A..29..581Y, 2005ARA&A..43..677S, 2013ARA&A..51..207B}. \citet{2014ApJ...787L..23L} found that the ${\rm R_{mid\ CO}}$ is largely independent of the FIR color (C(60/100)) in 65 luminous IR galaxies (LIRGs), and the defined ${\rm R_{mid\ CO}}$ can be used to distinguish the energy source of galaxies. According to the definition of ${\rm R_{mid\ CO}}$, we find that in NGC 1266, log ${\rm R_{mid\ CO}}= -$3.69 $\pm$ 0.10, while \citet{2014ApJ...787L..23L} found that log ${\rm R_{mid\ CO}}$ in SFGs is $-$4.13 $\pm$ 0.10 (an average value). This could be attributed to the fact that shocks can heat the gas more efficiently through a mechanical process, making a larger contribution to the mid- and high-\emph{J} CO emissions. Therefore, the ratio in NGC 1266 is greater than that in SFGs. By comparing the ${\rm R_{mid\ CO}}$ values, we obtain qualitative evidence that shocks may dominate the gas excitation in NGC 1266.

In addition, the CO SLED constructed with multiple CO lines is also one of the powerful tools for analyzing ISM (e.g., \citealt{2010A&A...518L..42V, 2013ApJ...762L..16M, 2013ApJ...779L..19P}). Subsequently, we modeled the CO observations by using existing models ({\tt RADEX}+{\tt shock}), and the results are shown in Figure \ref{Fig.2}. In this figure, the blue circles are the observations of CO, and the black circles are the best-fitting (reduced $\chi^2$ = 2.0) combination of the existing models. The three shaded areas represent: the error in the data (green), the optimal solution for the C-type shock (magenta) and RADEX (yellow). The best-fitting C-type shock has a hydrogen number density of $n_{\rm H}$ = $5 \times 10 ^4$ cm$^{-3}$ and velocity of $v_{\rm s}$ = $25 \pm 5$ km s$^{-1}$, and RADEX has a hydrogen number density of $n_{\rm H_2}$ = $1 \times 10 ^4$ cm$^{-3}$ and Kinetic temperature of $T_{\rm kin}$ = $30 \pm 5$ K.

The results show that the C-type shock can reproduce the CO emission lines of mid- and high-$J$ well ($J_{\rm up}$ = 4--10), while the low-$J$ CO emissions can be simulated by {\tt RADEX} model. Through the simulation of CO SLED, we confirmed the results of previous work that shock dominates the excitation of the gas in NGC 1266 \citep{2013ApJ...779L..19P}. This result is consistent with the above-mentioned diagnostic result using $R_{\rm mid\ CO}$. The shock velocity in our best-fitting parameters is consistent with previous work, while n$_{\rm H}$ is more than twice that of the previous work (\citealt{2013ApJ...779L..19P}). This is to be expected since the uncertainty in the shock velocity is a few km s$^{-1}$, and n$_{\rm H}$ only changes the final $\chi^2$ value \citep{2013ApJ...762L..16M}. For {\tt RADEX}, the result is consistent with the result of \citet{2015ApJ...800..105G}.

\subsection{Various methods to calculate SFR}
\label{section3.2} 
\begin{figure}[!hbtp]
   \centering
   \includegraphics[width=\textwidth, angle=0]{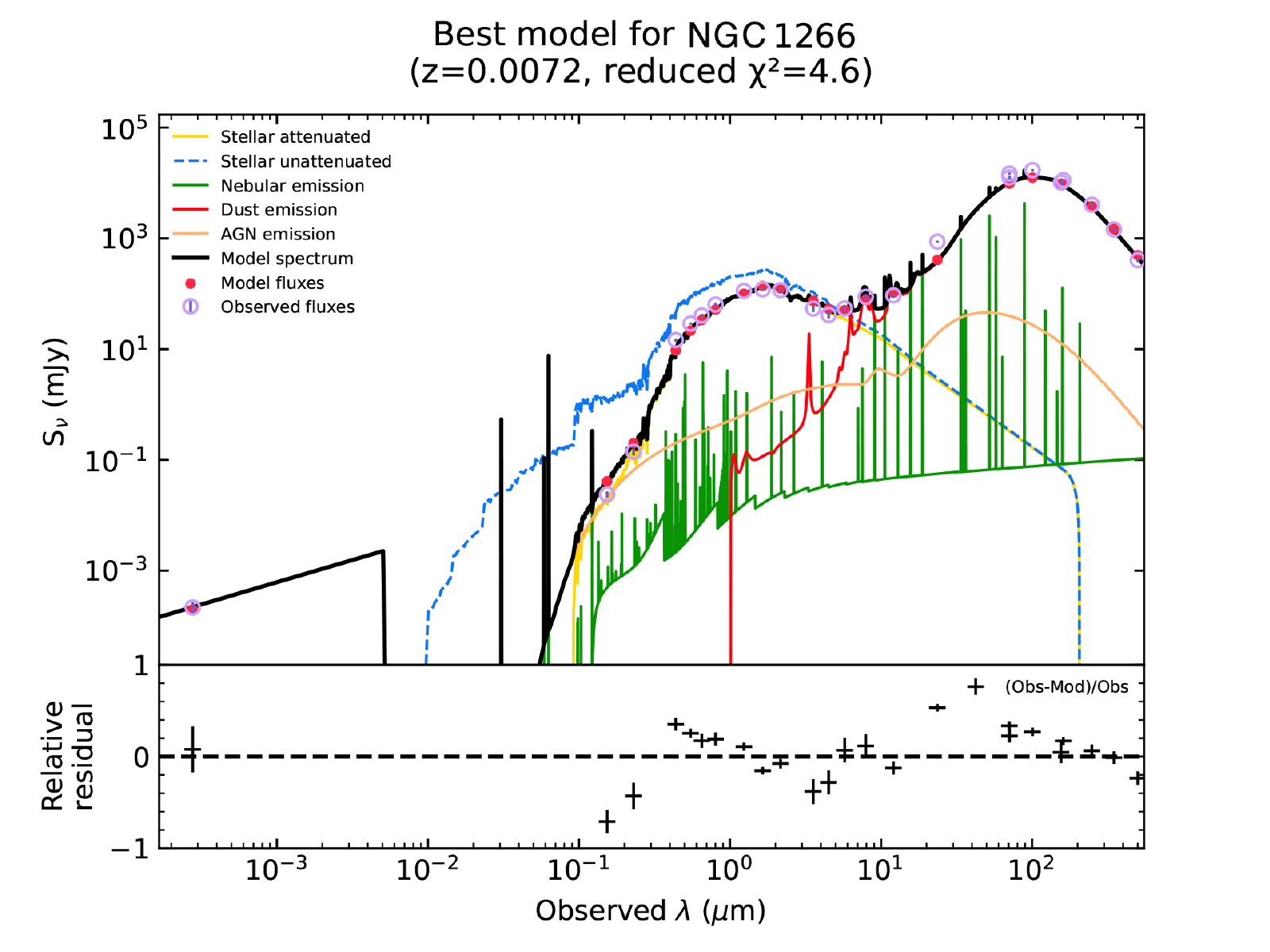}
   \caption{Panchromatic SED for NGC 1266 based on the photometry measurements listed in Table \ref{Table2} overlaid with the best-fitting SED model inferred from the SED fitting tool {\tt X-CIGALE} (black curve). The remaining informations are given in the legend of the figure. The panel at the bottom shows the comparison between the residuals of the model and the observed fluxes in each waveband.}
   \label{Fig.4}
   \end{figure}

There are a number of reasons that the calculation of the SFR in NGC 1266 is challenging. The nuclear region of NGC 1266 is highly obscured ($N_{\rm H} \thicksim 10^{24}$ cm$^{-2}$), and contains large amounts of shocked gas. \citet{2012MNRAS.426.1574D} indicated that the shock will contaminate the ionized gas emission lines, which hinders SFR calculation based on these lines from the H\,{\sc ii} regions. Furthermore, the hidden AGN \citep{2012MNRAS.426.1574D} has non-negligible contribution to MIR (10 $\rm{\mu m}$ in Figure \ref{Fig.4}, the AGN contribution is up to 10$\%$), and the ionized photons from the AGN also enhance the UV emission \citep{2015ApJ...798...31A}. The mid-\emph{J} CO emissions can also be used for the SFR calculation (e.g., \citealt{2015ApJ...810L..14L, 2015ApJ...802L..11L}), but the mid- and high-\emph{J} CO emissions are seriously affected by the shock (Fig \ref{Fig.2}), and the result must be severely overestimated (Table \ref{Table3}).

We construct the FUV-to-submm multi-wavelength SED, and perform fitting with the code {\tt X-CIGALE}. In Figure \ref{Fig.4}, the best-fitting SED from {\tt X-CIGALE} is overlaid on the observed data. In order to describe the bump clearly present in the optical band in the fitting process, the age of the main stellar population (SP) in the model is initialized to be 8--12 Gyr.
The best-fitting result compared to \citet{2019A&A...621A..51H} shows that considering this SP (the best matched age is 8 Gyr) can better match the optical observations. The derived SFR from the SED fitting is 1.17 $\pm$ 0.47 \emph{M$_{\odot}$}yr$^{-1}$. Furthermore, the AGN component has a derived bolometric luminosity ($L_{\rm bol}$) of 1.25$\times$10$^{42}$ erg s$^{-1}$, implying it is a low luminosity nucleus. It is worth noting that we also explored fitting the SED without the X-ray data and the measurement of SFR and the AGN luminosity were not affected, thanks to the constraint from high quality data from UV/optical to FIR. This is similar to the findings in \citet{2022A&A...658A..35K}, who used {\tt X-CIGALE} to study the host galaxy properties of a sample of X-ray AGN. The inclusion of hard X-ray data point is crucial to determine the contribution from AGN robustly in the longer wavelength range \citep{2020MNRAS.491..740Y}.   

We also utilize the [C\,{\sc ii}] 158 $\rm{\mu m}$ and [N\,{\sc ii}] 205 $\rm{\mu m}$ emission line measurements to further calculate the SFR.  The ionization potential of neutral carbon (11.3 eV) is lower than that of neutral hydrogen (13.6 eV), so the origins of [C\,{\sc ii}] may be multifold \citep{1991ApJ...373..423S}. Furthermore, when considering large IR luminosity, the emissivity of [C\,{\sc ii}] will decrease (e.g., \citealt{2012ApJ...755..171S, 2014A&A...568A..62D}). However, \citet{2019ApJ...886...60S} found that there was not deficit of [C\,{\sc ii}] in the neutral hydrogen regions. Therefore, the [C\,{\sc ii}] from the neutral regions can be used for SFR calibration regardless of the lower emissivity. Fortunately, [N\,{\sc ii}] 205 $\rm{\mu m}$ emission only arises from the H\,{\sc ii} regions, and its ratio to [C\,{\sc ii}] is almost constant, independent of the electron density ($n_{\rm e}$) \citep{2017ApJ...845...96C}. Meanwhile, their critical densities are similar, so [N\,{\sc ii}] 205 $\rm{\mu m}$ can be used to separate the [C\,{\sc ii}] emission originated from the neutral and ionized gas (e.g., \citealt{2006ApJ...652L.125O, 2009ApJ...691L...1W, 2014ApJ...782L..17D}).

The line flux ratio of [N\,{\sc ii}] 122 $\rm{\mu m}$ to [N\,{\sc ii}] 205 $\rm{\mu m}$ is an excellent probe for low-density ionized gas (e.g., \citealt{2011ApJS..195...12T, 2016ApJ...826..175H, 2017ApJ...845...96C}). When the electron temperature is $\sim 8000$ K, it is very sensitive to the variation of $n_{\rm e}$ (10 $\sim$ 300 cm$^{-3}$) (e.g., \citealt{2016ApJ...826..175H, 2017ApJ...845...96C}), enabling measurement of the $n_{\rm e}$. With the theoretical relationship between the line flux ratio of [C\,{\sc ii}] to [N\,{\sc ii}] 205 $\rm{\mu m}$ and $n_{\rm e}$ from \citet{2017ApJ...845...96C}, the line flux ratio of [C\,{\sc ii}] to [N\,{\sc ii}] 205 $\rm{\mu m}$ in the ionized region can be obtained. Finally, the proportion of [C\,{\sc ii}] emission from the H\,{\sc ii} regions can be estimated. The result shows that the $n_{\rm e}$ is 120 $\pm$ 6 cm$^{-3}$ in NGC 1266, and the corresponding ratio ([C\,{\sc ii}]/N\,{\sc ii}] 205 $\rm{\mu m}$) is 3.7 $\pm$ 0.2. That is, 12.0 $\pm$ 3.5 $\%$ of the total [C\,{\sc ii}] emission comes from the H\,{\sc ii} regions, and the remaining 88.0 $\pm$ 3.5 $\%$ comes from the neutral regions. As shown in Table \ref{Table3}, the SFR derived from the total [C\,{\sc ii}] 158 $\rm{\mu m}$ is 1.74 $\pm$ 0.18 \emph{M$_{\odot}$}yr$^{-1}$, whereas the SFR derived from the [C\,{\sc ii}] 158 $\rm{\mu m}$ that is associated with the neutral gas phase is 1.38 $\pm$ 0.14 \emph{M$_{\odot}$}yr$^{-1}$.

In addition, \citet{2013ApJ...765L..13Z} showed that there is a nearly linear relationship between the [N\,{\sc ii}] 205 $\rm{\mu m}$ luminosity and the total IR luminosity in galaxies, indicating that this emission line may be used for SFR tracing. The 60-to-100 $\rm{\mu m}$ flux density ratio observed by \emph{IRAS} for NGC 1266 is 0.78 ($f_{60}/f_{100}$), which is the ``warm" case as defined by \citet{2016ApJ...819...69Z}. The corresponding formula (listed in Table \ref{Table3}, and references therein) can be adopted for [N\,{\sc ii}] 205 $\rm{\mu m}$ calculation, which leads to a derived SFR as 3.87 $\sim$ 6.46 $M_{\odot}$yr$^{-1}$. There are two groups of different fitting coefficients for the relationship between [N\,{\sc ii}] 205 $\rm{\mu m}$ and SFR \citep{2016ApJ...819...69Z}, so this result is a range.

Also, using the mid- and high-$J$ CO emissions \citep{2015ApJ...810L..14L, 2015ApJ...802L..11L}, we can trace SFR, especially CO(7-6). The result shows that the derivation result must be overestimated (listed in Table \ref{Table3}) because the CO emissions are severely affected by the shock (Figure \ref{Fig.2}).

\begin{table}
\begin{center}
\caption{Independently derived SFR for NGC 1266}\label{Table3}
\begin{threeparttable}
 \begin{tabular}{cccccc}
  \hline\hline\noalign{\smallskip}
Spectral &  \begin{tabular}[c]{@{}l@{}}Luminosity \\ ($L_{\odot}$)\end{tabular}      & Calculation formulas & \begin{tabular}[c]{@{}l@{}}SFR\tnote{a}\\ (\emph{M$_{\odot}$}yr$^{-1}$)\end{tabular}                    \\
  \hline\noalign{\smallskip}
IR       & 1.30 $\pm$ 0.39 $\times$ 10$^{10}$                                                      & SFR(\emph{M$_{\odot}$}yr$^{-1}$) = 1.73 $\times$ 10$^{-10}$ $L_{\rm IR}$($L_{\odot}$)\tnote{b}               & 2.25 $\pm$ 0.68      \\

[C\,{\sc ii}] 158$\rm{\mu m}$      & 0.30 $\pm$ 0.03 $\times$ 10$^{8}$                                                     & log SFR(\emph{M$_{\odot}$}yr$^{-1}$) = a log $L_{\rm [C_{II}]}$(erg s$^{-1}$) $-$ b\tnote{c}               & 1.74 $\pm$ 0.18                   \\

[N\,{\sc ii}] 205$\rm{\mu m}$      & 0.28 $\pm$ 0.09  $\times$ 10$^{7}$                                                     & log SFR(\emph{M$_{\odot}$}yr$^{-1}$) = a log $L_{\rm [N_{II}]}$(L$_{\odot}$) - b\tnote{d}               & 3.87 $\sim$ 6.46                   \\
CO(7-6)       & 1.36 $\pm$ 0.15 $\times$ 10$^6$                                                     & SFR(\emph{M$_{\odot}$}yr$^{-1}$) = 1.31 $\times$ 10$^{-5.00 \pm 0.12}$($L_{\rm CO(7-6)}$/$L_{\odot}$)\tnote{e}               & $\sim$ 17.82                  \\
X-CIGALE       & --                                                     & --               & 1.17 $\pm$ 0.47                  \\
  \noalign{\smallskip}\hline
\end{tabular}
\begin{tablenotes}
 \footnotesize
 \item[a]  Values calculated directly from the calibration formulas.  The results derived by the others are summarized in Table \ref{Table4}. 
  \item[b] \citet{1998ARA&A..36..189K}. 
  \item[c] From \citet{2019ApJ...886...60S}. For the total emission, a and b are 1.02, $-$41.64; For neutral emission, a and b are 0.98, $-$40.05.
  \item[d] From \citet{2016ApJ...819...69Z}, when the $f_{60}/f_{100}$ of the target source is between 0.6$\sim$0.9 (warm). For this formula, there are different fitting coefficients: a is 1, b is 5.64, or a is 0.98, b is 5.53. The SFR obtained in the two cases is 6.64 $M_{\odot}$yr$^{-1}$ and 3.87 $M_{\odot}$yr$^{-1}$ respectively.
  \item[e] \citet{2015ApJ...802L..11L}, CO lines are seriously affected by shock, so we do not give error in the calculation result.
\end{tablenotes}
\end{threeparttable}
\end{center}
\end{table}

\begin{table}
    \centering
    \caption{Summary of SFR for NGC 1266 derived in the literature}
    \begin{threeparttable}
    \begin{tabular}{ccc}
    \hline\hline
    Methods & \begin{tabular}[c]{@{}l@{}}SFR\\
    (\emph{M$_{\odot}$}yr$^{-1}$)\end{tabular} & References\tnote{d} \\ \hline
    PAH+FUV & 0.3  &1 \\ 

    [Ne\,{\sc ii}]  & 1.5   &  1 \\ 
    IR\tnote{a}     & 2.2   & 1 \\
    CIGALE model    & 1.46 $\pm$ 1.12  & 2 \\ 
    GRASIL model    & 1.40 $\pm$ 1.20   & 2 \\ 
    MAGPHYS model   & 0.45 $\pm$ 1.15\tnote{c}  &  2 \\ 
    FUV+TIR\tnote{b}         & 1.79  & 2  \\ 
    H$_{\alpha}$ + 24 $\rm{\mu m}$ & 1.98   & 2 \\ 
    TIR\tnote{b} & 5.37 $\pm$ 1.23   &  3  \\
    \hline
    \end{tabular}

    \begin{tablenotes}
    \footnotesize
    \item[a]  Assuming that IR is all from SF, $L_{\rm IR}$ is calculated using the IRAS four--band fluxes.
    \item[b] $L_{\rm TIR}$ is the total infrared-submm energy budget of galaxies, its range is 8--1000 $\rm{\mu m}$.
    \item[c] The reason for the large error is that the model does not consider the AGN component.
    \item[d] 1 is \citet{2015ApJ...798...31A}, 2 is \citet{2019A&A...621A..51H}, 3 is \citet{2014A&A...568A..62D}
    \end{tablenotes}
    \end{threeparttable}
\label{Table4}

\end{table}

\section{DISCUSSION} \label{section4}
NGC 1266 has been investigated previously (e.g., \citealt{2011ApJ...735...88A, 2015ApJ...798...31A, 2015ApJ...800..105G, 2019A&A...621A..51H}). Combining the existing observations (e.g., \emph{CARMA, ALMA, XMM-Newton}), \cite{2015ApJ...798...31A} not only gave the range of SFR in NGC 1266, but also corrected the mass loss rate reported previously \citep{2011ApJ...735...88A}, and discussed the possible nuclear activity. \citet{2015ApJ...800..105G} studied the characterization of nuclear molecular gas in NGC 1266 combined with \emph{Herschel} and literature archival data. By fitting to the CO lines, they gave strong evidence that the warm outflow gas is driven by AGN and excited by shock, because the SF is too weak. Also, \citet{2013ApJ...779L..19P} used \emph{Herschel} and the ground-based observational CO data to obtain evidence that shocks dominate the gas excitation with the help of existing models (e.g., PDR and Shock).

\subsection{Cross-check of SFR derivation} \label{sub1}
The derived SFR using the [C\,{\sc ii}] 158 $\rm{\mu m}$ associated with neutral regions is 1.38 $\pm$ 0.14 \emph{M$_{\odot}$}yr$^{-1}$, and the derivation result of the total [C\,{\sc ii}] 158 $\rm{\mu m}$ emissions is 1.74 $\pm$ 0.18 \emph{M$_{\odot}$}yr$^{-1}$. Compared to the SFR given by the SED fitting ($\sim$1.2 \emph{M$_{\odot}$}yr$^{-1}$) and the values reported in the literature (Table \ref{Table4}), these are in good agreement, especially considering the scatter of the empirical formula (listed in Table \ref{Table3}, and references therein).

Obviously, the derived SFR from the measured [N\,{\sc ii}] 205 $\rm{\mu m}$ flux is significantly larger, three times more than that of [C\,{\sc ii}] 158 $\rm{\mu m}$. For the relationship between [N\,{\sc ii}] 205 $\rm{\mu m}$ and SFR, \citet{2016ApJ...819...69Z} fitted each sample using a least-squares, geometrical mean functional relationship with a linear form (listed in Table \ref{Table3}). The formula listed in Table \ref{Table3} has two groups of fitting coefficients according to the definition of \citet{2016ApJ...819...69Z}, the SFR obtained is 6.46 $M_{\odot}$yr$^{-1}$ and 3.87 $M_{\odot}$yr$^{-1}$ respectively, and the scatter is 0.25 dex. We suggest that the reason is because the dependence of [N\,{\sc ii}] 205 $\rm{\mu m}$ on FIR color ($f_{60}/f_{100}$) has the significant contribution to the total scatter of the [N\,{\sc ii}] 205 $\rm{\mu m}$-SFR relation \citep{2016ApJ...819...69Z}. For mid-/high-$J$ CO emissions, taking CO(6-5) as an example, the shock contributes up to $\sim$ 76$\%$ to its emissions, which is also shown in Figure \ref{Fig.2}. Moreover, the ratio of CO luminosity simulated by the shock model with the best parameters to the IR luminosity is 0.46 $\times$ 10$^{-3}$ ($L_{\rm CO}/L_{\rm IR}$), which is very close to the observed ratio (0.54 $\times$ 10$^{-3}$). Therefore, shocks can well stimulate the mid-/high-$J$ CO emission to the observed level. In fact, the result of CO SLED analysis shows that the mid-\emph{J} CO transitions are severely affected by the shocks (Fig \ref{Fig.2}), and consequently its calculation result must be overestimated (Table \ref{Table3}). If the presence of the AGN and shocks in NGC 1266 is fully ignored, and all the IR flux is attributed to SF, the derived SFR from IR (2.25 $\pm$ 0.68 $M_{\odot}$yr$^{-1}$) can be viewed as the upper limit. This is consistent with the conclusions in the literature (e.g., \citealt{2014A&A...568A..62D}). 

As shown in Table \ref{Table4}, previous work have used various methods (emission lines or existing models) to calculate the SFR of NGC 1266. \citet{2015ApJ...798...31A} suggested that due to the combined effect of AGN and shock in NGC 1266, the derived SFR (about 0.3 $M_{\odot}$yr$^{-1}$) using PAHs + UV should be the lower limit of its SFR. Moreover, we can find that all the results are all around 2.0 $M_{\odot}$yr$^{-1}$ in tables \ref{Table3} and \ref{Table4}, regardless of the tracer used (e.g., continuums or emission lines). So we think that the SFR range in NGC 1266 may be between 0.3 $M_{\odot}$yr$^{-1}$ -- 2.0 $M_{\odot}$yr$^{-1}$. Although the derived values of some methods are obviously large in these two tables, as mentioned in Section \ref{section3.2}: UV, MIR and H\,{\sc ii} regions emission lines are affected by AGN in different levels, and shock will affect the PAHs and gas emission lines, so these results are expected \citep{2012MNRAS.426.1574D, 2015ApJ...798...31A}. In our work, we perform calculations with [C\,{\sc ii}] from the neutral hydrogen regions after correcting for deficit. Comparing the results in Table 3 and combining our SED fitting result, we suggest that the [C\,{\sc ii}] calculation result may be more reliable. 

\subsection{Evidence for a low luminosity AGN} \label{sub2}
Although it is postulated that the AGN in NGC 1266 has strong mechanical energy to drive the massive molecular outflow \citep{2011ApJ...735...88A,2014ApJ...780..186A,2013ApJ...779..173N}, it is inconclusive whether the compact continuum source resolved by \emph{ALMA} is dominated by an AGN or an ultra-compact starburst \citep{2015ApJ...798...31A}. The observations of dense gas tracers (e.g., CS(2-1) and HCN(1-0)) by \emph{ALMA} and \emph{CARMA} had shown that its nuclear region lies behind a large gas column density ($N_{\rm H_2} \thicksim 3\times 10^{24}$ cm$^{-2}$, or $N_{\rm H} \thicksim 6\times 10^{24}$ cm$^{-2}$, \citealt{2015ApJ...798...31A}). Such a large gas column density is considered to be Compton-thick, which can obscure an intrinsically bright AGN as well as attenuate X-rays. 

In one of the scenarios, \citet{2015ApJ...798...31A} assumed that the compact FIR continuum source (within 30 pc of the nucleus) is heated by an AGN, the IR emission associated with AGN will account for 66$\%$ of the total flux density. Their spectral modeling using {\em Chandra} and {\em XMM}-Newton indicates the absorption corrected 2--10 keV luminosity of the putative AGN is $7\times 10^{43}$ erg s$^{-1}$. However, \citet{2008ARA&A..46..475H} indicated that the bolometric luminosity of nearby galactic nuclei span $\thicksim$10$^{38}$ -- 10$^{44}$ erg s$^{-1}$, with a median value of $L_{\rm bol}$ = 3 $\times$ 10$^{40}$ erg s$^{-1}$ and half of the nearby galactic nuclei lying between 3 $\times$ 10$^{39}$ and 3 $\times$ 10$^{41}$ erg s$^{-1}$. Therefore, the 2--10 keV luminosity of the putative AGN obtained by \citet{2015ApJ...798...31A} exceeds the IR luminosity in the nucleus. Noting that the column density is poorly constrained due to limited energy range of {\em Chandra} and {\em XMM}-Newton, and the hard X-ray may include scattered and reflected components, this is suggested to be the upper limit of the hard X-ray luminosity \citep{2015ApJ...798...31A}. Alternatively, scaling from the bolometric luminosity of the central source, they further estimate an AGN with 2--10 keV luminosity $\sim 10^{42}$ erg s$^{-1}$ obscured by $N_H=3\times 10^{24}$ cm$^{-2}$ can match the observed X-ray flux.

In contrast, according to our best-fit multi-wavelength SED (AGN emission in Figure \ref{Fig.4}, yellow curve), it implies that the fraction of IR luminosity from AGN is lower than the assumption above. The derived AGN bolometric luminosity $L_{\rm bol}$ is $1.25\times$10$^{42}$ erg s$^{-1}$, so its $L_{\rm bol}/L_{\rm X}$ = 27.80, which is very close to the value ($L_{\rm bol}/L_{\rm X}$ = 28.0) given by \cite{2008ARA&A..46..475H}. Assuming $N_{\rm H}=6\times 10^{24}$ cm$^{-2}$, with an estimated intrinsic 2--10 keV luminosity of $10^{42}$ erg s$^{-1}$ \citep{2015ApJ...798...31A}, one would expect to find a \emph{NuSTAR} count rate (2.6 $\pm$ 0.6 $\times$ 10$^{-2}$) at least one order of magnitude higher than the observed (1.8 $\pm$ 0.6 $\times$ 10$^{-3}$). Therefore the hidden AGN is a low luminosity one, with an Eddington ratio of $L_{\rm bol}/L_{\rm Edd}=6\times 10^{-3}$ (for a black hole mass $M_{\rm BH}=1.7\times 10^6$M$_{\odot}$, \citealt{2015ApJ...798...31A}). The presence of a radio jet \citep{2013ApJ...779..173N} is also consistent with the theoretical regime for the low luminosity AGN \citep{2021NatAs...5..928S} and Seyfert where hot accretion flow becomes important \citep{2014ARA&A..52..529Y, 2019MNRAS.490.2567M}.  We conclude that it is more likely the central compact emission is dominated by starburst in NGC 1266.

\section{CONCLUSIONS}\label{section5}
Comparing with previous work: we adopt the combination of {\tt SHOCK+RADEX}. Although the final results are not significantly different. Furthermore, we also confirm this result qualitatively ($R_{\rm mid\ CO}$ value). For the construction of galaxy SED, we use {\tt X-CIGALE} which contains more AGN templates (e.g., {\tt CIGALE}), and we consider different SPs to describe the bump clearly present in the optical band in the fitting process. Most importantly, we include the X-ray data in the fitting process. In addition, we also calibrate the galaxy SFR using different methods, especially the [C\,{\sc ii}] only from the neutral regions.

In this paper, we present our results of the inner activity in a S0 galaxy, NGC 1266. With the help of existing models as well as the R$_{\rm mid CO}$ value, we confirm the mechanism responsible for gas excitation in the system. Due to the special case that NGC 1266 exists, we provide different SFR calibration results, especially [C\,{\sc ii}] from the neutral regions. The most difference is that we include X-ray data from \emph{NuSTAR} in the SED fitting process and use it to differentiate the activity inside the galaxy. The main conclusions of our work are as follows:

(1) The ${\rm R_{mid\ CO}}$ value in NGC 1266 is larger than that in SFGs, which is consistent with the result of shock heating. The fitting results of CO SLED also show that shocks dominate the gas excitation in the galaxy. 

(2) The SED of the galaxy is constructed and modeled by the {\tt X-CIGALE}, and a series of physical parameters are obtained, indicating that the SFR is 1.17 $\pm$ 0.47 $M_{\odot}$yr$^{-1}$. In the fitting we consider different parameter settings to better describe the bump clearly present in the optical band. Meanwhile, the [C\,{\sc ii}] only from neutral regions also show consistent result 1.38 $\pm$ 0.14 $M_{\odot}$yr$^{-1}$. The higher  measurement from [N\,{\sc ii}] 205 $\rm{\mu m}$ is affected by the uncertainty of empirical calibration and presence of shocks. We also provide the related results from others in Table \ref{Table4}.

(3) Based on our fitted multi-wavelength SED, we find that the compact emission related to AGN does not account for 66$\%$ of the total flux density as previously expected. If the gas column density is indeed $\thicksim 6\times 10^{24}$ cm$^{-2}$, the hidden Compton-thick AGN could be detected by hard X-ray observations that are sensitive to $>10$~keV.  We used the 3--79 keV archival data from \emph{NuSTAR} hard X-ray observation to constrain this and found a marginal 3$\sigma$ detection. An obscured low luminosity AGN is more favorable. Therefore, the central compact emission is probably dominated by starburst in NGC 1266.

\begin{acknowledgements}


We are grateful to the anonymous referee for her/his thoughtful review and very constructive suggestions, which greatly improved this paper. We thank Dr. Guang Yang for helpful instruction on using X-CIGALE. This work is supported by the National Natural Science Foundation of China  (NSFC grant Nos. 12033004, U1831205, 12173079 and 12221003). J.W. acknowledges the science research grants from the China Manned Space Project with No. CMS-CSST-2021-B02 and CMS-CSST-2021-A06. This work is based in part on observations made with \emph{Herschel}, an European Space Agency Cornerstone Mission with significant participation by the National Aeronautics and Space Administration (NASA). This research has made use of the NASA/IPAC Extra-galactic Database (NED), operated by the Jet Propulsion Laboratory of the California Institute of Technology, under contract with NASA.
\end{acknowledgements}

\appendix\label{APP}
\section{The {\tt X-CIGALE} module usage and parameters setting}\label{appA}
{\bf SFH}: it contains several SFH forms, such as {\tt sfh2exp}, {\tt sfh$\_$buat08} and {\tt sfhdelayed} \citep{2019A&A...622A.103B}. One of the most popular is ``delayed" SFH modeling, however its most obvious limitation is that it does not allow for recent quenching of the SFR. To address this issue, \citet{2017A&A...608A..41C} extended SFH to allow for instantaneous resent change in SFR, and called it module {\tt sfhdelayedbq}. \citet{2019A&A...621A..51H} has successfully modeled the KINGFISH galaxies using this SFH module, so here we also adopt this SFH module: {\tt sfhdelayedbq}.

\begin{table}[]
\centering
\caption{Parameters setting of each module in code X-CIGALE}
\begin{threeparttable}
\begin{tabular}{lll}
\hline\hline
Modules\tnote{a}                       & Parameters\tnote{b}         & Values                                                    \\ \hline
sfhdelayedbq & tau$\_$main        & 500,1000,2000,4000,8000 (Myr)                             \\ \cline{2-3} 
                              & age$\_$main\tnote{c}        & 8000,10000,12000 (Myr)                                    \\ \cline{2-3} 
                              & age$\_$bq          & 10,100,500,1000 (Myr)                                     \\ \cline{2-3} 
                              & r$\_$sfr           & 0.01,0.05,0.1,0.5,1.0,5.0,10.0                            \\ \cline{2-3} 
                              & sfr$\_$A           & 1.0 ($M_{\odot}$yr$^{-1}$)                                                       \\ \hline
bc03         & imf                & 1                                                         \\ \cline{2-3} 
                              & metallicity        & 0.02 ($M_{\odot}$)                                                      \\ \cline{2-3} 
                              & separation$\_$age  & 10 (Myr)                                                  \\ \hline
nebular                       & logU               & $-2.0$                                                    \\ \hline
distant$\_$modified$\_$CF00   & A$_{\rm V}\_$ISM   & 0.1,0.5,1.0,1.5,2.0 (mag)                                       \\ \hline

dl2007\tnote{d}       & qpah               & 0.47,1.12,2.5,3.9,4.58                                    \\ \cline{2-3} 
                              & umin               & 0.10,0.15,0.30,0.50,0.80,1.00,1.20,1.50,2.00,4.00,10.,25. (Habing) \\ \cline{2-3} 
                              & umax               & 1e6 (Habing)                                                       \\ \cline{2-3} 
                              & gamma              & 0.001,0.003,0.009,0.01,0.03,0.09,0.1                      \\ \hline
fritz2006    & r                  & 60.0                                                      \\ \cline{2-3} 
                              & $\tau$             & 1.0                                                       \\ \cline{2-3} 
                              & $\beta$            & -0.5                                                      \\ \cline{2-3} 
                              & $\gamma$           & 4.0                                                       \\ \cline{2-3} 
                              & $\theta$           & 100.0 (\textdegree)                                                     \\ \cline{2-3} 
                              & $\Psi$             & 50.1 (\textdegree)                                                      \\ \cline{2-3} 
                              & $f_{\rm AGN}$      & 0.01,0.05,0.1,0.2,0.3,0.5 ($\%$)                                 \\ \cline{2-3} 
                              & E(B-V)             & 0.04,0.1                                                  \\ \hline
xray                          & $\Gamma$           & 1.8                                                       \\ \hline

\end{tabular}
\begin{tablenotes}
 \footnotesize
 \item[a] In addition to the templates listed, there is a final module called {\tt redshifting} during setup, which first redshifts the spectrum and dims it, multiplying the wavelength by 1+\emph{z} and dividing the spectrum by 1+\emph{z}. Also, it considers the shorter wavelength radiation absorbed by the IGM \citep{2019A&A...622A.103B}.
 \item[b] Parameters not listed in the module are set to default values
 \item[c] Age of the main stellar population in a galaxy \citep{2019A&A...622A.103B}
 \item[d] \citet{2007ApJ...657..810D}
\end{tablenotes}
\end{threeparttable}
\label{paraset}
\end{table}

{\bf SSPs}: it is calculated using the \citet{2003MNRAS.344.1000B} SSPs, where the metallicity \emph{z} is 0.02 and the Initial Mass Function is from \citet{2003PASP..115..763C};

{\bf Nebular emission}: the {\tt CLOUDY} model is used in the model to calculate the templates, and they have the same metallicity and stellar population (SP). The ionization parameter in this module uses the model default values ($log$U = $-$2.0), and assumes that the fraction of Lyman continuum photons escaping the galaxy is zero, i.e. the photon does not directly heat the dust \citep{2019A&A...621A..51H}.

{\bf Dust attenuation}: the model adopts the modified starburst attenuation law to explain the absorption of stellar and nebular radiation by interstellar dust, which takes into account the reddening differences of SPs of different ages \citep{2019A&A...622A.103B}. Here, we adopt the \citet{2000ApJ...539..718C} attenuation law.

{\bf Dust emission}: the model includes several dust emission templates, which assuming that the dust emission is optically thin \citep{2007ApJ...657..810D, 2014ApJ...784...83D, 2014ApJ...780..172D}. Also, it considers the possible variations of the polycyclic aromatic hydrocarbon (PAH) and radiation field intensity ($U_{\rm min}$ and $U_{\rm max}$), as well as the fraction illuminated from $U_{\rm min}$ to $U_{\rm max}$.

{\bf AGN}: As discussed above, the model contains two different AGN templates -- smooth \citep{2006MNRAS.366..767F} and clumpy \citep{2012MNRAS.420.2756S, 2016MNRAS.458.2288S}. They take into account three components through a radiative transfer model: a source located in the torus, the scattered emission by dust, and the thermal dust emission \citep{2019A&A...622A.103B}. They are set by seven parameters: r the ratio of the inner and outer radii of the dust torus, $\tau$ the optical depth at 9.7 $\rm{\mu m}$, $\gamma$ and $\beta$ the horizontal/vertical density distribution of the dust, $\theta$ the aperture angle of the dust torus, $\Psi$ the angle between the AGN axis and the line-of-sight direction, and $f_{\rm AGN}$ the AGN fraction. In our run, we adopt the AGN templates from \citet{2006MNRAS.366..767F}.

{\bf X-ray}: {\tt X-CIGALE} uses a new module, X-ray, which enables the addition of X-ray data during the fitting process. Input data requirements are absorption-corrected \citep{2020MNRAS.491..740Y}, but do not distinguish absorption from the source itself, the Milky Way, or the intergalactic medium (IGM).




\end{document}